\documentstyle[psfig,twocolumn,aps]{revtex}
\begin{document}

\draft
\wideabs{
\title{Spin {\it versus} Lattice Polaron:
Prediction for Electron-Doped CaMnO$_3$}

\author{Yiing-Rei Chen and Philip B. Allen}
\address{Department of Physics and Astronomy, State University of New York,
Stony Brook, New York 11794-3800}

\date{\today}

\maketitle

\begin{abstract}
CaMnO$_3$ is a simple bi-partite antiferromagnet(AF) which can
be continuously electron-doped up to LaMnO$_3$. Electrons enter
the doubly degenerate $E_{g}$ subshell with spins aligned
to the $S=3/2$ core of Mn$^{4+}(T_{2g}^{3\uparrow})$. We take
the Hubbard and Hund energies to be effectively infinite.  Our model
Hamiltonian has two $E_{g}$ orbitals per Mn atom, nearest
neighbor hopping, nearest neighbor exchange coupling of
the $S=3/2$ cores, and electron-phonon coupling of Mn orbitals
to adjacent oxygen atoms.  We solve this model for
light doping.  Electrons are confined
in local ferromagnetic(FM) regions (spin polarons) where there
proceeds an interesting competition between spin polarization
(spin polarons) which enlarges the polaron, and lattice polarization
(Jahn-Teller polarons) which makes it smaller.
A symmetric 7-atom ferromagnetic cluster (Mn$_7^{27+}$)
is the stable result, with net spin $S$=2 relative to the undoped AF.
The distorted oxygen positions around the electron are predicted.
The model also predicts a critical doping
$x\simeq 0.045$ where the polaronic insulator becomes unstable
relative to a FM metal.       
\end{abstract}
}

\section{Introduction}

CaMnO$_3$ is a bi-partite (G-type) antiferromagnetic(AF)
insulator\cite{Wollan} with N\'eel temperature $T_N$=125K,
and almost perfect cubic perovskite crystal structure.
There is not a large literature on this material.
It deserves attention as a particularly simple case of an
AF insulator which can be electron-doped.
The $(T,x)$ phase diagram \cite{Jonker1,Jonker2} of the
Ca$_{1-x}$La$_x$MnO$_3$ series has attracted 
attention because of the fascinating
interplay of spin order, orbital order, and metallic {\it versus}
insulating transport.
``Colossal magnetoresistance'' (CMR) occuring at
concentration $x\approx 0.65$ and $T\approx 250$ K
is the most dramatic manifestation \cite{CMR}.
For small $x$, magnetization and conductivity measurements
\cite{Neum,Maignan} suggest
local ferromagnetic(FM) regions or ``spin polarons''
in the range $0.02<x<0.06$.
In this paper we use a model for pure CaMnO$_3$, the $x=0$ end
member, and predict its behavior under light doping, $x \ll 1$.
We keep $T$ equal to 0 and neglect lattice zero-point energy,
but all other degrees
of freedom are allowed and interact in interesting ways.

Our main question is, what is
the ground state of an excess electron in
CaMnO$_3$?  The answer to this question will also apply to lightly
doped CaMnO$_3$ provided the doped state is homogeneous.
Our model has four parameters. (1) The bandwidth parameter $t$
governs the effective Mn $E_{g}$ electron
hopping between Mn$^{3+}$ and Mn$^{4+}$ through the intervening
oxygen.  The spins of the two Mn ions must be parallel as in the
usual ``double exchange'' model\cite{Zener,Anderson}.
(2) The magnetic exchange parameter $J$ couples spins of first-neighbor
Mn ions, due to virtual hopping of $T_{2g}$ electrons.
(3) The electron-phonon coupling constant $g$ describes
interactions between Mn $E_{g}$ orbitals and the 6
nearest oxygen atoms.  (4) Oxygen displacements $u$
are opposed by the restoring force $-Ku$, where $\omega=\sqrt{K/M}$
is an Einstein frequency assigned to oxygen vibrations along the bonds.
There are two important dimensionless parameters.
Spin polarons \cite{Nagaev} are controlled by the ratio
$\beta=t/JS^2$.  Jahn-Teller (JT) lattice polarons \cite{Thomas}
are controlled by the parameter $\Gamma=g^2/Kt$
\cite{Allen}.  Balancing these competing effects, we find the most
favorable local FM spin arrangement, lattice distortion and
electron wavefunction.  As doping increases, we
predict a transition from polaronic insulator to FM metal.

\section{Model Hamiltonian}

The Mn$^{4+}$ ion in CaMnO$_3$ has configuration $3d^3$,
{\it i.e.}, the three spin-aligned $T_{2g}$ states 
$(xy,yz,zx)$ are filled with
electrons, while the two spin-aligned $E_{g}$ states
$(\psi_2=(x^2-y^2)/\sqrt{3}, \psi_3=3z^2-r^2)$
are empty and lie above by the crystal field splitting. The empty
opposite-spin $T_{2g}$ states are split to even higher energy
by the Hund term $J_H$.  Light electron doping puts carriers into the
doubly-degenerate $E_{g}$ level. Hopping of
$(dd\pi)$-type occurs from $T_{2g}$ to $T_{2g}$,
and of $(dd\sigma)$-type from $E_{g}$ to $E_{g}$,
but no $E_{g}$ to $T_{2g}$ hopping matrix element exists
because of the simple cubic structure.
Virtual $T_{2g}$ hopping \cite{Goodenough} (at the cost
of Hubbard energy $U$) gains delocalization energy 
if adjacent spins are antiparallel.
This gives an $S=3/2$ antiferromagnetic 
Heisenberg Hamiltonian with exchange coupling
$J=2(dd\pi)^2/U$ \cite{Fradkin}, $U$ being large compared
with $(dd\pi)$ \cite{Satpathy},
and agrees with the experimentally
observed magnetic structure of pure CaMnO$_3$.
The ground state has $\uparrow$ spins on the
A sublattice (when $\exp(i\vec{Q}\cdot\vec{l})=1$,
where $\vec l$ labels the Mn sites), and $\downarrow$
spins on the B sublattice (when $\exp(i\vec{Q}\cdot\vec{l})=-1$)
with $\vec{Q}=(\pi,\pi,\pi)$.

\begin{figure}[htbp]
\centerline{\psfig{figure=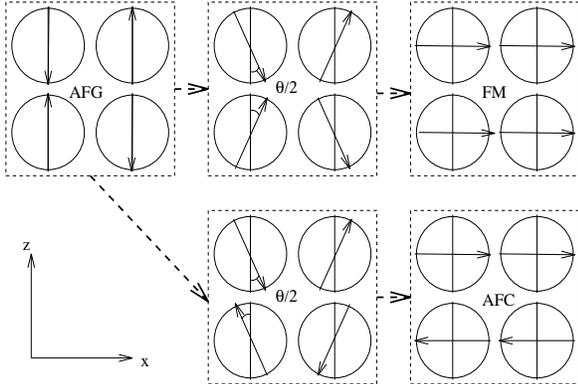,height=2.0in,width=3.0in,angle=0}}
\caption{Schematic spin structures for the antiferromagnetic G (AFG),
ferromagnetic (FM), and antiferromagnetic C (AFC) structures, and
interpolating canted structures.}
\label{fig:spinangle}
\end{figure}

Ignoring for now the electron-phonon terms, the Hamiltonian
for an excess electron is
${\cal H}={\cal H}_t+{\cal H}_J$. The first term contains
hopping of Mn $E_{g}$ electrons to nearest neighbors,
\begin{eqnarray}
{\cal H}_t&=&t\sum_{l,\pm}[S(\vec{l},\vec{l}\pm\hat{z})
c_3^{\dagger}(\vec{l}\pm\hat{z}) c_3(\vec{l}) \nonumber \\
&+&\mbox{rotations to $\hat{x}$,$\hat{y}$ directions}],
\label{eq:hopping}
\end{eqnarray}
\begin{equation}
S(1,2)=\cos\frac{\theta_1}{2} \cos\frac{\theta_2}{2}
	+\sin\frac{\theta_1}{2} \sin\frac{\theta_2}{2}
	e^{-i(\phi_1 -\phi_2)}
\label{eq:spinfactor}
\end{equation}
Here $c_3(\vec{l})$ destroys an electron state $\psi_3(\vec{r}-\vec{l})$
on the Mn atom at $\vec{l}$;
$\vec{l}\pm \hat{z}$ labels the Mn neighbors above and below the
one at site $\vec{l}$; and $t$ is the $(dd\sigma)$ integral from
Slater-Koster two-center theory \cite{Slater}.  We
use a value $t=-0.75$ eV obtained from fitting the band structure
of CaMnO$_3$ \cite{Pickett,Dagotto}(see Sec. III).  The factor $S(1,2)$
comes from locally rotating the axis of spin quantization
for the $i^{\rm th} \ E_{g}$ electron into the direction
$(\theta_i,\phi_i)$ of the $i^{\rm th} \ S=3/2$ core spin,
treating the angles $(\theta_i,\phi_i)$ as classical parameters,
and discarding from the Hilbert space the state with spin opposite
to the core spin ({\it i.e.} assuming
$J_H\rightarrow\infty$ \cite{Dagotto,Choi}).
In this paper, we usually take the spins to be perfectly ordered at $T=0$,
that is, $\theta=\theta_1-\theta_2$ equals 0 or $\pi$,
corresponding to $S(1,2)$
equal to 1 or 0 for FM or AF oriented neighbors.
We will also consider uniformly canted states where a relative angle
$\theta=\pi-\theta_0$ occurs, with $\phi$=constant, so that $S$
takes the value $\sin(\theta_0 /2)$.  These spin orientations are shown
in Fig. \ref{fig:spinangle}.

The rotation of $\psi_3$ to the
$\hat{x}$ axis is $(-\psi_3+\sqrt 3\psi_2)/2=3x^2-r^2$, and to the
$\hat{y}$ axis it is $(-\psi_3-\sqrt 3\psi_2)/2=3y^2-r^2$.
Using these, we rewrite ${\cal H}_t$ in the usual orthonormal basis
($\psi_2,\psi_3$),
\begin{equation}
{\cal H}_t=t\sum_{l,\delta =x,y,z}
\left(
\begin{array}{cc}
c_2^{\dagger} (\vec{l})&
c_3^{\dagger} (\vec{l}) \end{array}
\right){\sf T}_{\delta}\left(
\begin{array}{c}
c_2 (\vec{l}\pm \hat{\delta} ) \\
c_3 (\vec{l}\pm \hat{\delta} ) \end{array} \right),
\label{eqn:Ht}
\end{equation}
where it is understood that the hopping only operates between
parallel spin Mn atoms.  The hopping matrices are
\begin{equation}
{\sf T}_x=\left(
\begin{array}{cc}
\frac34 & -\frac{\sqrt 3}4\\
-\frac{\sqrt 3}4 & \frac14 \end{array} \right), \ \
{\sf T}_y=\left(
\begin{array}{cc}
\frac34 & \frac{\sqrt 3}4\\
\frac{\sqrt 3}4 & \frac14 \end{array} \right), \ \
{\sf T}_z=\left(
\begin{array}{cc}
0 & 0\\
0 & 1 \end{array} \right).
\end{equation}

The ${\cal H}_J$ term is the AF nearest neighbor $T_{2g}$
exchange:
\begin{equation}
{\cal H}_J=\sum_{<l,l^{\prime}>}
J\vec {S}(\vec{l})\cdot\vec{S}(\vec{l}^{\prime})
\label{eq:HJ}
\end{equation}
The exchange coupling $J$ is can be estimated from Mean Field theory
(we use a quantum treatment for spin 3/2; FM and AFM answers are equal)
to be $JS^2=3.23$ meV, using the measured N\'eel temperature
$T_{\rm N}=125$ K \cite{Zeng}.
However, for a given coupling $J$, a more accurate
estimate from susceptibility expansions reduces the $T_{\rm C}$
of the Heisenberg ferromagnet\cite{Wood1}
\begin{equation}
\frac {T_{\rm C}}{T_{\rm C(MF)}}
=\frac {\frac {5}{96}(z-1)[11S(S+1)-1]}{\frac {2}{3}zS(S+1)},
\end{equation}
where $z=6$ is the number of nearest neighbors.
For a Heisenberg antiferromagnet\cite{Wood2,Fisher} it
is estimated that $T_N$ is slightly higher,
\begin{equation}
\frac {T_{\rm N}}{T_{\rm C}}\simeq 1+\frac{0.63}{zS(S+1)}.
\end{equation}
Making these corrections, the coupling is found to be $JS^2=4.74$ meV,
giving $t/JS^2$=158, with probable uncertainty of 10\%.  The phonon
parts of the Hamiltonian are given in Sec. V.

\section{uniform solutions}

First consider the hypothetical case of a uniform FM spin
order.  Then a doped-in electron could hop
without paying a Hund penalty, and extended Bloch states
would form with wavefunctions
\begin{equation}
\Psi_k= d_2\frac1{\sqrt{N}}
\sum_l e^{i\vec{k}\cdot\vec{l}} \psi_2(\vec{l})+
d_3\frac1{\sqrt{N}} \sum_le^{i\vec{k}\cdot\vec{l}} \psi_3(\vec{l}),
\end{equation}
where $d_2,d_3$ are coefficients.  Diagonalizing
Eq.(\ref{eqn:Ht}), the resulting energy eigenvalue,
taking into consideration that $t$ is negative, is
\begin{eqnarray}
\frac{E({\vec k})}{|t|}&=&
-\cos k_x-\cos k_y-\cos k_z \nonumber\\
& &\mp
(\cos^2k_x+\cos^2k_y+\cos^2k_z \nonumber\\
& &-\cos k_x\cos k_y-\cos k_y\cos k_z-\cos k_z\cos k_x
)^{1/2}.\nonumber\\
\label{eq:Ek}
\end{eqnarray}
At $\cos k_x=\cos k_y=\cos k_z=1$, the energy is
minimum, $E(\vec{k}=0)=-3|t|$, the state being doubly
degenerate.  Although the actual spin arrangement is not FM, the result
$|3t|$ sets a useful scale for the maximal
energy gain from electron delocalization.  Also, the dispersion relation
(\ref{eq:Ek}) can be fitted to a FM CaMnO$_3$ majority-spin band
structure calculated in density-functional theory
\cite{Pickett}, determining the hopping energy $t=-0.75$ eV.

For light doping, the magnetic energy Eqn.(\ref{eq:HJ}) of
antiferromagnetic order ($-zJNS^2/2$ for classical spins;
$N$ is the number of Mn ions) is
much larger than hopping energy.  As discussed 
by deGennes \cite{deGennes}, the best uniform solution 
(for a one-band model) is a compromise
where antiferromagnetic spins cant uniformly toward FM solutions.
Start from the N\'eel AF structure and let all spins
tilt towards the $\hat{x}$ direction with angle $\theta/2$,
as shown in Fig. \ref{fig:spinangle}.
This costs magnetic energy $\Delta E_J=zNJS^2(1-\cos\theta)/2$.
A doped-in electron can now delocalize, reducing its energy by
$-3|t|\sin(\theta/2)$.  For small doping $x=n/N \ll 1$,
the total hopping energy is $\Delta E_t=-3n|t|\sin(\theta/2)$.  Therefore when
$\sin(\theta/2)=x|t|/4JS^2$, the total energy
$\Delta E=\Delta E_J+\Delta E_t$ is minimum,
$-3 Nx^2t^2/8JS^2$.  From this estimate, the system cants all the way
to FM when the optimal $\theta$ equals $\pi$,
which occurs at $x=4JS^2/|t|\simeq 0.025$.

\begin{figure}[htbp]
\centerline{\psfig{figure=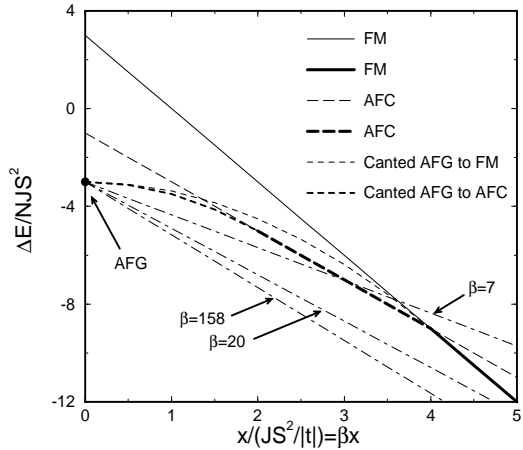,height=2.5in,width=3.0in,angle=0}}
\caption{Energy per electron of various doped states.
Electron concentration $x$ is measured in unitsof $1/\beta \equiv JS^2/|t|$.
Bold lines denote the most favorable uniform states.  Non-uniform
(polaron) states (discussed in Sec. VI; using $\Gamma$=0.25)
are denoted by dot-dashed lines
for the various values of $\beta$ indicated.
Among different extended
states, AFC lies lowest from $\beta x=2\sim 4$, and FM takes over above
$\beta x=4$ ($x\simeq 0.025$). For light doping, the inhomogeneous polaron
states are stable. As $x$ increases, uniform states take over.
The transition from polaron-doped AFG to FM occurs at $x\simeq 0.045$
(using $\Gamma =0.25$, $t=0.75$ eV, $JS^2=4.74$ meV, $\beta\simeq 158$).
Notice that the value of $\beta$ determines the critical doping $x_c$
of the phase transition, and even the transition type, as small
$\beta$ results in AFG to AFC transition, rather than to FM.
But for small $\beta$ the predicted $x_c$ can be large
enough to exceed the light-doping approximation used here.}
\label{fig:FGC}
\end{figure}

Because the doubly-degenerate
$E_{g}$ electrons hop anisotropically, a better canted 
solution exists.  Let all spins with $\exp(i\vec{K}\cdot\vec{\l})=1$
($\vec{K}$ is $(0,\pi,\pi)$) tilt towards
$+\hat x$ and all spins with $\exp(i\vec{K}\cdot\vec{\l})=-1$ tilt
towards $-\hat x$
with angle $\theta/2$, as shown in Fig. \ref{fig:spinangle}.
The magnetic energy cost is reduced to
$\Delta E_J=zNJS^2(1-\cos\theta)/6$, but the delocalizaton energy
lowering is also reduced to $\Delta E_t=-2n|t| \sin(\theta/2)$.
This type of canting terminates in
C-type antiferromagnetism when $x=2JS^2/t$, as shown in Fig.
\ref{fig:spinangle}.  Fig. \ref{fig:FGC} summarizes
our results for the energies of uniform solutions
{\it versus} $x$.

\section{Pure Spin Polaron Effect}

Experimentally, CaMnO$_3$ is insulating at small doping $x$.
This rules out uniform solutions with electrons doped into delocalized states.
The simplest picture is that each doped electron is localized on one Mn site,
creating a local Mn$^{3+}$ with spin $S=2$ in an AF
background, for a spin excess of 1/2.
This does not agree with the measured saturated
magnetization \cite{Neum}, which has been interpreted in terms of
local spin flipping (net excess spin of 2) in the region $0.02<x<0.08$.
Local spin flipping leads to local FM regions (called
spin polarons) around doping centers, allowing the
doped electron to gain delocalization energy.  In this section we discuss
candidate localized ground states of a single $E_{g}$ electron,
using the same Hamiltonian ${\cal H}={\cal H}_t+{\cal H}_J$.

First we make a continuum (effective mass)
approximation in the spirit of Nagaev \cite{Nagaev}.
Inside the local FM region, electrons hop like
free electrons with inverse band mass $1/m^{\ast}=(a/\hbar)^2|t|$
given by Eqn.(\ref{eq:Ek}), with $a$ the perovskite lattice constant,
$a\simeq 3.73\AA$.
This electron sits in a spherical well with infinite walls at radius $R$,
whose depth is $-3|t|$.
G-type antiferromagnetism resumes for $r>R$.
The ground state energy of the electron in this polarized region is then
\begin{equation}
\Delta E_t=|t|(-3+\frac{\pi^2 a^2}{2R^2}),
\end{equation}
where the second term is the zero-point energy in the well.
The magnetic energy cost from spin-canting is
\begin{equation}
\Delta E_J=zJS^2(4\pi R^3/3a^3),
\end{equation}
where the last factor is the number of Mn atoms in the cluster.
Optimizing $\Delta E=\Delta E_t+\Delta E_J$ over R,
the optimal size cluster has a number
of Mn atoms inside the sphere equal to
\begin{equation}
\frac 43\pi \left( \frac{R}{a} \right)^3
=\frac 43\pi\left(\frac{\pi|t|}{4zJS^2}\right)^{3/5}
               \simeq 26.
\end{equation}
This is close to the optimum 25-site symmetric cluster
which we will find by exact diagonalization and show in Table I.
However, we will also find a smaller asymmetric cluster of lower
energy.

It is interesting that continuum theory gives a lower energy
if the region $r<R$ is canted rather than fully FM.
The A sublattice is fixed for all $r$, but the B sublattice
is tilted toward the A sublattice by angle $\theta$ for $r<R$.
Then the optimum tilting angle is 180$^{\circ}$ (FM) for 
$\beta=t/JS^2>\beta_c=(7\pi^2/18)^{-5/2}4z/\pi \simeq 220$, 
while for smaller $\beta$, the optimal tilting angle
is $\sin(\theta/2)=\beta/\beta_c$. 
Our estimated value is $\beta\simeq 158$, which gives
$\beta/\beta_c \simeq 0.72$. So the spins inside radius $R$
are approximately 90$^{\circ}$ apart, and the optimum cluster
increases to 31.

Now we repeat the calculation using the true discrete Hamiltonian.
First consider flipping only one spin. In this way, the spin-flipped Mn
(the central site), along with its 6 nearest neighbors, form a 7-site
cluster with all 7 spins parallel.  The cluster is invariant under
transformations of the point group $O_h$. If the central Mn spin is
kept unflipped, but instead the spins of those 6 nearest neighbors are
flipped, a 25-site $O_h$-symmetric cluster is formed. Similar steps
can be taken to obtain larger and larger symmetric clusters.
We will not look at candidate states with canted spins except
for one special case to be mentioned later.

Each spin flip costs magnetic energy $6\times 2JS^2$.  
Since spins are parallel inside the cluster, the electron 
can hop among the $E_{g}$ orbitals of all the spin-aligned 
Mn ions, with a corresponding energy lowering from 
delocalization. Table I shows numerical values of ground state 
energy found by exact diagonalization of $2M \times 2M$ 
Hamiltonian matrices for symmetric clusters of $M$ Mn atoms, 
ranging in size from $M$=1 to 63.  The ground states are all 
doubly degenerate (E$_{g}$ representation) because 
of the $O_h$ symmetry of the cluster.

\begin{figure}[htbp]
\centerline{\psfig{figure=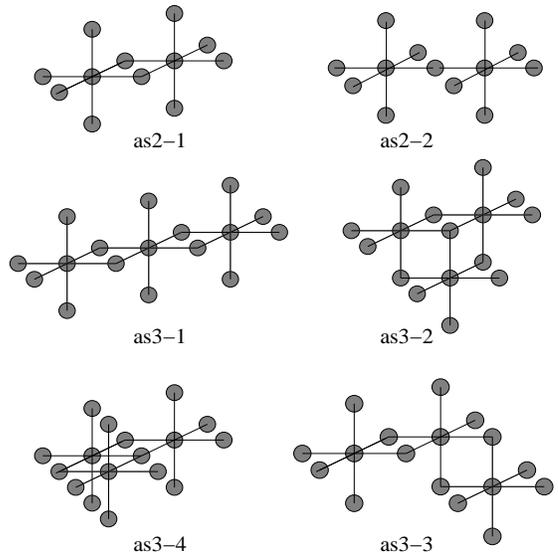,height=3.0in,width=3.0in,angle=0}}
\caption{Different asymmetric clusters: as2 is the only one with two  
spins flipped; there are four types of as3 which have 3 spins
flipped.}
\label{fig:5as}
\end{figure}

Asymmetric clusters are also possible. If we flip one spin at site
$\vec{l}$ and another one at site $\vec{l}+\hat{x}+\hat{y}$, a 12-site
cluster is formed. Starting from this 12-site cluster, there are four
different ways to create a larger cluster with one more flipped
spin, as shown in Fig. \ref{fig:5as}. Other possibilities are less
closely packed, such as the 13-site cluster shown in the figure,
and other 3-spin-flipped cases not shown here. The ground state
energy of these examples are calculated, as shown in Table II.
For our chosen values of $t$ and $JS^2$, the most favorable spin
polaron is an asymmetric 17-site cluster with three flipped spins.

\begin{table}
\begin{center}
\caption{Ground state energy of symmetric clusters, with only spin
polaron effects included. The last column uses $t=-0.75$ eV
and $JS^2=4.74$ meV. }
\begin{tabular}{cccc}
\makebox[1cm]{cluster} &
\makebox[2cm]{number of} &
\makebox[3cm]{energy gain from} &
\makebox[2cm]{ground state}\\
size & spins flipped & hopping $(|t|)$
& energy (eV)\\ \hline
1&0&0&0\\
7&1&$-\sqrt{3}=-1.732$&-1.242\\
25&6&-2.330 &-1.406\\
51&13&-2.449 &-1.097\\
57&14&-2.380 &-0.989\\
63&19&-2.600 &-0.869\\
\end{tabular}
\end{center}
\end{table}
\begin{table}
\begin{center}
\caption{Ground state energy of asymmetric clusters, with pure spin
polaron effect.}
\begin{tabular}{lccc}
\makebox[1.3cm]{cluster} &
\makebox[2cm]{number of} &
\makebox[2.8cm]{energy gain from} &
\makebox[1.9cm]{ground state}\\
\multicolumn{1}{c}{size} & spins flipped & hopping $(|t|)$
& energy (eV)\\ \hline
12(as2-1)&2&-1.936 &-1.338\\
13(as2-2)&2&-2     &-1.386\\
16(as3-2)&3&-2.015 &-1.341\\
17(as3-1)&3&-1.936 &-1.281\\
17(as3-3)&3&-1.984 &-1.317\\
17(as3-4)&3&-2.145 &-1.438\\
\end{tabular}
\end{center}
\end{table}

There are many possible variations with inhomogeneously canted spins, of which
we considered only one, a 25-site cluster with the 7 inner atoms
canted rather than ferromagnetically aligned with the outer 18 atoms.
This interpolates between the 7 and the 25 atom cluster.  Without
canting, the 7 and 25 atom cluster become equal in energy for the
value $|t|/JS^2\simeq 100$, smaller than our preferred value of 158.
It turned out that in the range $94<|t|/JS^2<104$, the locally
canted state was lower in energy than either the 7 or the 25 atom
pure ferromagnetic cluster.

\section{Lattice Polaron Effect}

The degenerate ground state, found in the previous section for
symmetrical spin polarons, is Jahn-Teller (JT)
unstable\cite{JT}. We now add to our Hamiltonian lattice
distortions, controlled by the electron-phonon interaction.  The
only lattice degrees of freedom included are oxygen motions
along the bonds to the nearest two Mn ions.
If an oxygen moves along this bond a distance $u$,
it gets closer to one Mn atom
and farther from another.  This will raise by $gu$ (for the closer Mn)
or lower by $gu$ (for the farther Mn) the energy
of any occupied Mn $E_{g}$ state of the type which points toward
the oxygen (that is, the $3z^2-r^2$ state if oxygen motion in the
$\hat{z}$-direction is considered.)
We use adiabatic approximation (oxygen mass $\to \infty$) and
treat the oxygen distortions as classical parameters.  Each Mn ion is
surrounded by six oxygens whose distortion
amplitudes $(u_{\delta}(\vec{l}\pm\hat{\delta}/2),
\hat{\delta}=\hat{x},\hat{y},\hat{z})$
form basis vectors for a representation of $O_h$,
namely, $a_{1g}\oplus e_{g}\oplus t_{1u}$. The vector
representation $t_{1u}$ contains the distortions
$u_z(\vec{l}+\hat{z}/2)+u_z(\vec{l}-\hat{z}/2)$,
etc.  The remaining 3 degrees of freedom
$(Q_z =u_{z}(\vec{l}+\hat{z}/2)
-u_{z}(\vec{l}-\hat{z}/2), etc.)$
form basis vectors of
$a_{1g}\ (Q_1=(Q_x+Q_y+Q_z)/\sqrt{6})$ and
$e_{g}\ (Q_2=(Q_x-Q_y)/2,
Q_3=(2Q_z-Q_x-Q_y)/\sqrt{12})$, in Van Vleck
notation\cite{VanV}.  The oxygen distortions are limited by a harmonic
restoring force, {\it e.g.}, $-Ku_z(\vec{l}\pm \hat{z}/2)$.

The lattice elastic energy and electron-phonon
interaction terms of the Hamiltonian are \cite{note}
\begin{eqnarray}
{\cal H}_{\rm L}&=& \frac K2\sum_l[
u_x(\vec{l}+\frac 12\hat{x})^2+u_y(\vec{l}+\frac 12\hat{y})^2+
u_z(\vec{l}+\frac 12\hat{z})^2],\nonumber\\
& &\\
{\cal H}_{\rm ep}&=&-\frac{4g}{\sqrt 3}\sum_l[
c_z^{\dagger}(\vec{l})c_z(\vec{l})Q_z(\vec{l})\nonumber\\
& &\hspace{2cm}+\mbox{rotations to $\hat{x}$, $\hat{y}$
directions}].
\label{eq:Hep}
\end{eqnarray}
${\cal H}_{\rm ep}$ can be split into two parts, JT and ``breathing'':
\begin{equation}
{\cal H}_{\rm ep}={\cal H}_{\rm JT}+{\cal H}_{\rm br}
\end{equation}
where, in Van Vleck notation,
\begin{eqnarray}
{\cal H}_{\rm JT}&=&2g\sum_l
(\begin{array}{cc} c_2^{\dagger}(\vec{l})&c_3^{\dagger}(\vec{l}) \end{array})
\left(
\begin{array}{cc}
Q_3(\vec{l})&Q_2(\vec{l})\\
Q_2(\vec{l})&-Q_3(\vec{l}) \end{array} \right)
\left(
\begin{array}{c}
c_2(\vec{l})\\
c_3(\vec{l}) 
\end{array}   
\right)\nonumber\\
\label{eq:HJT}
{\cal H}_{\rm br}&=&-2\sqrt{2}g\sum_l
Q_1(\vec{l})\left(
c_2^{\dagger}(\vec{l})c_2(\vec{l})+
c_3^{\dagger}(\vec{l})c_3(\vec{l}) \right).
\label{eq:Hbr}
\end{eqnarray}
The vector distortions ($t_{1u}$) do not couple to
$E_{g}$ electron states and therefore do not appear.

A simple case shows how ${\cal H}_{\rm JT}$ splits energy
degeneracy. Suppose a doped electron is localized at a
single Mn site, with no hopping or spin flipping
considered. The two $E_{g}$ states at that Mn are the only
degrees of freedom for the electron and are originally
degenerate. When the six surrounding oxygen distortions
are considered, the
degeneracy is lifted in ${\cal H}_{\rm JT}$:
\begin{eqnarray}
{\cal H}_{\rm JT}&=&2g(
\begin{array}{cc}
c_2^{\dagger}&c_3^{\dagger}
\end{array}
)\left(
\begin{array}{cc}
Q_3&Q_2\\
Q_2&-Q_3
\end{array}
\right)\left(
\begin{array}{c}
c_2\\
c_3
\end{array}
\right)\\
&=&2g
\left(
\begin{array}{cc}
\alpha^{\dagger}&\beta^{\dagger}
\end{array}
\right)
\left(
\begin{array}{cc}
-Q&0\\
0&Q
\end{array}
\right)
\left(
\begin{array}{c}
\alpha\\
\beta
\end{array}
\right)
\end{eqnarray}
where
\[
\left(
\begin{array}{c}
\alpha^{\dagger}\\
\beta^{\dagger}
\end{array}
\right)=\left(
\begin{array}{cc}
\cos\frac{\phi}2 &-\sin\frac{\phi}2\\
\sin\frac{\phi}2 &\cos\frac{\phi}2
\end{array}
\right) \left(
\begin{array}{c}
c_2^{\dagger}\\
c_3^{\dagger}
\end{array}
\right)
\]
and $(Q_2,Q_3)=Q(\sin\phi,\cos\phi)$. The
change in energy due to ${\cal H}_{\rm JT}+{\cal H}_{\rm br}+{\cal
H}_{\rm L}$ is $\Delta E=-2\sqrt 2gQ_1\pm 2gQ +12(Q_1^2+Q^2)/2$.
The energy splitting $\pm 2gQ$ comes only from the JT term.
The optimal distortions are $Q_1=4\sqrt2 g/K$ and $Q=4g/K$,
which give the maximum energy lowering of ground state
$\Delta E=-6g^2/K$.

The angle $\phi$ does not enter $\Delta E$. This continuous
degeneracy can be lifted either by adding kinetic energy
and quantization of lattice degrees of freedom (dynamic
JT effect), or else by introducing higher order
anharmonic terms in ${\cal H}_{\rm ep}$\cite{Kanamori}
\cite{Millis}. However, later in the discussion of the 7-site
cluster, we shall show how this continuous degeneracy is
naturally removed by the increase of electronic and
lattice degrees of freedom.

It is interesting to consider what would happen if spins were
ferromagnetically ordered, so that magnetism does not assist
localization. Then polaron formation can only occur through
lattice distortion and is prohibited when the delocalization
energy per electron is larger than the polaron energy, {\it i.e.},
when $\Gamma=g^2/K|t|$ is less than a critical value close to 0.5.
We believe that $\Gamma$ is close to 0.25, so that in the hypothetical
ferromagnetic case, polarons would not form.  Antiferromagnetic
confinement is needed before a lattice polaron effect occurs.
LaMnO$_3$ is different in this respect; its cooperative Jahn-Teller
ground state makes polaron formation easier \cite{Allen}.

\section{The 7-site cluster: Spin and Lattice Polaron}

With the full Hamiltonian considered, the 7-site cluster
turns out to be the most important one, as shown later
in this section.
Only a single spin is flipped.  Inhomogeneous
canting will not be favored for this state.
By understanding its ground state algebraically,
we learn several interesting aspects of the influence of
${\cal H}_{\rm ep}$ in the Hamiltonian.
The dimensionless parameter
$\Gamma\equiv g^2/Kt$ characterizing
the strength of electron-phonon coupling has a value
near 0.25 for LaMnO$_3$, and we assume that the
value for for CaMnO$_3$ is similar, {\it i.e.}
$0.20<\Gamma <0.30$\cite{Oki}\cite{Popovic}. We will
measure energies in units of $|t|$, using dimensionless
lattice variables $Q^{\prime}\equiv \sqrt{K/|t|}\,Q$.
The prime in $Q^{\prime}$ is suppressed from here on.

To study the ground state of a single electron in
the 7-site cluster, a 14-dimensional Hilbert space is
used, consisting of atomic
$E_{g}$ orbitals from each of the 7 Mn atoms.
Later when we turn on ${\cal H}_{\rm ep}$, we will find
that for $\Gamma $ small, fewer than 14
basis functions are needed for the ground state calculation.
The 14-dimensional space can be decomposed into 7 irreducible
representations of point group $O_h$, namely,
$A_{1g}\oplus A_{2g}
\oplus T_{1u}\oplus T_{2u}\oplus 3E_{g}$.
When lattice distortions are absent
$( \Gamma =0,{\cal H}_{\rm ep}={\cal H}_{\rm L}=0)$,
these functions diagonalize the 14-dimensional ${\cal H}_t$.
The 3 $E_{g}$-type basis functions
$( \psi _2^{+},\psi _3^{+})
$, $( \psi_2^0,\psi _3^0) $ and
$( \psi _2^{-},\psi _3^{-}) $ are
degenerate separately with eigenvalues $+\sqrt{3}|t|$,
$0$, and $-\sqrt{3}|t|$. These 6 states, along
with the $A_{1g}$ state (with eigenvalue 0), will be the main
states of interest when $\Gamma \neq 0$.
All other states stay absent as long as $\Gamma $ is small.

When $\Gamma \neq 0$, ${\cal H}_{\rm ep}$ and ${\cal H}_{\rm L}$
are turned on. Amplitudes of lattice distortions appear linearly
in the matrix representation of ${\cal H}_{\rm ep}$,
and quadratically in that of ${\cal H}_{\rm L}$. Since these
lattice distortions are of order $\sqrt{\Gamma }$
when $\Gamma \ll 1$, ${\cal H}_{\rm ep}$ and ${\cal H}_{\rm L}$
can be treated as perturbations to ${\cal H}_t$ . In the
following, all lattice distortion modes will
be introduced.  Then second order perturbation of the original
ground states $(\psi _2^{-},\psi _3^{-})$ can be expressed in
terms of these modes and shows that most of these modes do not
participate in the ground state lattice distortion. This will
in turn eliminate the need for considering electron states of
symmetries which couple to the absent modes only. Finally,
from the form of the perturbed ground state energy, the pattern
of its optimized lattice distortion will be derived and compared
to the exact numerical result.

There are 36 oxygens adjacent to one or more Mn atoms 
in the 7-atom cluster, so there are 36 distortion
parameters in ${\cal H}_{\rm ep}$. These can
be organized into sets of basis vectors for irreducible
representations of the point group $O_h$, namely,
$3a_{{1g}}\oplus a_{{2g}}\oplus 5t_{{\rm 1u}}
\oplus 3t_{{\rm 2u}}\oplus 4e_{g}$. 
The $a_{{1g}}$-type modes are denoted as $(q_{i1}$,
where $i=1,2,3)$, and the $e_{g}$-type modes are denoted as
$(q_{i2},q_{i3})$, where $i=1,2,3,4$. The matrix elements of
${\cal H}_{\rm ep}$ can be reexpressed in terms of these modes,
and ${\cal H}_{\rm L}$ is simply $\sum_{\rm modes}
q_{\rm modes}^2 /2$.

Second order perturbation theory shows that for small $\Gamma$,
many of these modes appear only quadratically with positive
coefficients and hence should be optimized to $0$.  For larger $\Gamma$,
these coefficients start to turn negative, and
the corresponding distortions start to develop.  The critical
values are $\Gamma=0.443$ for the appearance of the $t_{1u}$ 
and $q_{43}$ distortion, $\Gamma =1.30$
for $t_{2u}$, $\Gamma =0.819$ for $a_{2g}$ and $q_{42}$.  It will thus
happen that as $\Gamma$ increases, the symmetry
of the ground state electron wavefunction and lattice distortion
pattern is gradually lowered by the successive appearance of these
distortions.

We therefore ignore the above modes for the actual range
$\Gamma \approx 0.25 \pm 0.05$. This eliminates the
presence of electron states of $A_{2g}$, $T_{1u}$ 
and $T_{2u}$ symmetry. The remaining
modes are 3 sets of $(q_{i1},q_{i2},q_{i3}) $, with $i=1,2,3$,
of the inner, intermediate and outer oxygen layers, respectively.
Degenerate first-order perturbation theory for $(\psi_2^{-},\psi_3^{-})$ 
shows that the ground state energy of ${\cal H}_t$ splits into
\begin{eqnarray}
\frac{E^{(1)}}{\left| t\right| } &=&-\sqrt{3}-\sqrt{\Gamma }(\sqrt{2}Q_1\mp
\sqrt{Q_2^2+Q_3^2}) \\
Q_1&\equiv &\frac 13q_{11}+\frac 16q_{21}+\frac 1{12}q_{31},\nonumber\\
Q_2&\equiv &\frac 13q_{12}+\frac 16q_{22}+\frac 1{12}q_{31},\nonumber\\
Q_3&\equiv &\frac 13q_{13}+\frac 16q_{23}+\frac 1{12}q_{33}.\nonumber
\end{eqnarray}
Thus at first order, there is still a continuous degeneracy, in the
sense that the energy JT splitting depends only on
$Q\equiv \sqrt{Q_2^2+Q_3^2}$, not on $\phi \equiv
\tan^{-1} (Q_3/Q_2)$.

\begin{figure}[htbp]
\centerline{\psfig{figure=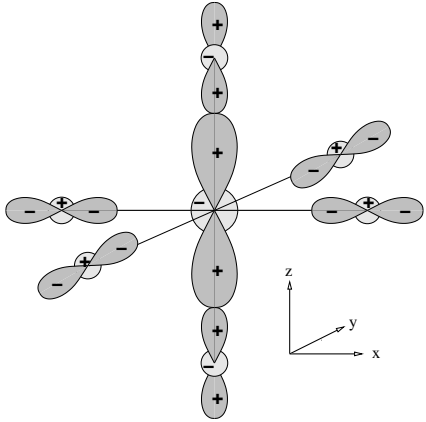,height=2.3in,width=2.3in,angle=0}}
\caption{One of the possible ground state wavefunctions in the 7-site
cluster. The corresponding lattice distortion pattern has $q_1$ and 
$q_3$ components. }
\label{fig:7Mn}
\end{figure}

To include 2nd order perturbations, for simplicity, we treat analytically
only the
distortion modes from the inner oxygen layer, $(q_{11},q_{12},q_{13})$.
The number of related electronic states is
now reduced to 5, namely, they are $(A_{1g},\psi _2^{+},\psi
_3^{+})$ and $(\psi _2^{-},\psi _3^{-})$. In this
5-dimensional subspace, the Schr\"odinger equation to be solved can be
expressed as follows (not including ${\cal H}_{\rm L}$, which is always
proportional to the identity matrix):
\begin{equation}
\left(
\begin{array}{cc}
{\cal H}_t^{II}+{\cal H}_{\rm ep}^{II}-E & h_{\rm ep} \\
h_{\rm ep}^T & {\cal H}_t^{I}+{\cal H}_{\rm ep}^{I}-E
\end{array}
\right) \left(
\begin{array}{c}
\Psi ^{II} \\
\Psi ^{I}
\end{array}
\right) =0
\end{equation}
where $\Psi ^{I}\equiv ( \psi _2^{-},\psi _3^{-}) $, $\Psi
^{II}\equiv (A_{1g},\psi _2^{+},\psi _3^{+})$. We obtain an effective
Hamiltonian ${\cal H}_{\rm eff}={\cal H}_t^{I}+{\cal H}_{\rm ep}^{I}
-h_{\rm ep}^T({\cal H}_t^{II}+{\cal H}_{\rm ep}^{II}-E)^{-1}h_{\rm ep}$ for
$\Psi ^{I}$. Since $E$ should be very close to $-\sqrt{3}| t| $
for small perturbation, we take $E$ as $-\sqrt{3}| t|\times
{\bf 1}_{(3\times 3)}$ and ignore ${\cal H}_{\rm ep}^{II}$
in the denominator. The ${\cal H}_{\rm eff}$
obtained this way shows that the degeneracy of $(\psi _2^{-},\psi
_3^{-})$ is now lifted to become (including ${\cal H}_{\rm L}$)
\begin{eqnarray}
\frac{E^{(2)}}{\left| t\right| } &=&-\sqrt{3}-\sqrt{2}A-\left( \frac{4\sqrt{3%
}}3-\frac 9{2\Gamma }\right) \left( A^2+\rho ^2\right) \pm \sqrt{\epsilon}
\nonumber\\
\epsilon &\equiv &\frac 43\rho ^4+\frac{32}3A^2\rho ^2+\frac{8\sqrt{6}}3%
A\rho ^2+\rho ^2\nonumber\\
& &\hspace{2cm}+\frac 43\rho ^3\left( 4\sqrt{2}A+\sqrt{3}\right) \cos
\left( 3\theta \right),
\end{eqnarray}
where we introduce the notation
\begin{eqnarray}
A &\equiv &\frac{\sqrt{\Gamma }}3q_{11},
\hspace{0.5cm} \rho \equiv \sqrt{B^2+C^2}\nonumber\\
B &\equiv &\frac{\sqrt{\Gamma }}3q_{12}\equiv \rho \sin \theta ,\:
C\equiv \frac{\sqrt{\Gamma }}3q_{13}\equiv
\rho \cos \theta  \nonumber
\end{eqnarray}
We see that $\cos 3\theta $ should be $+1$ to minimize the ground
state energy. Hence the degeneracy of ground states is not continuous
anymore and has become 3-fold. This feature agrees with the numerical
result. The ground state electronic wavefunction of $\theta =0$ is
shown in Fig. \ref{fig:7Mn}.
A rotation which brings $\hat{z}$ to $\hat{x}$
$(\hat{y}) $ will generate the wavefunctions of
$\theta =2\pi/3 $ $(\theta =4\pi/3)$. 

\section{Numerical Results}

\begin{figure}[htbp]
\centerline{\psfig{figure=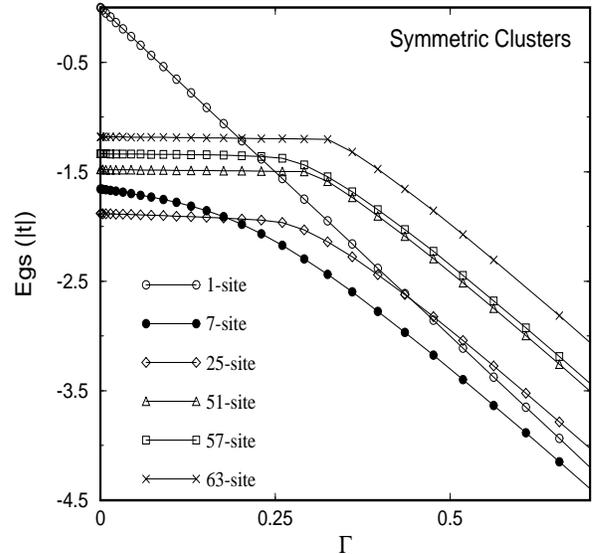,height=3.4in,width=3.4in,angle=0}}
\caption{Numerical results for ground state energy($E_{gs}$) of symmetric
clusters}
\label{fig:syf2}
\end{figure}

\begin{figure}[htb]
\centerline{\psfig{figure=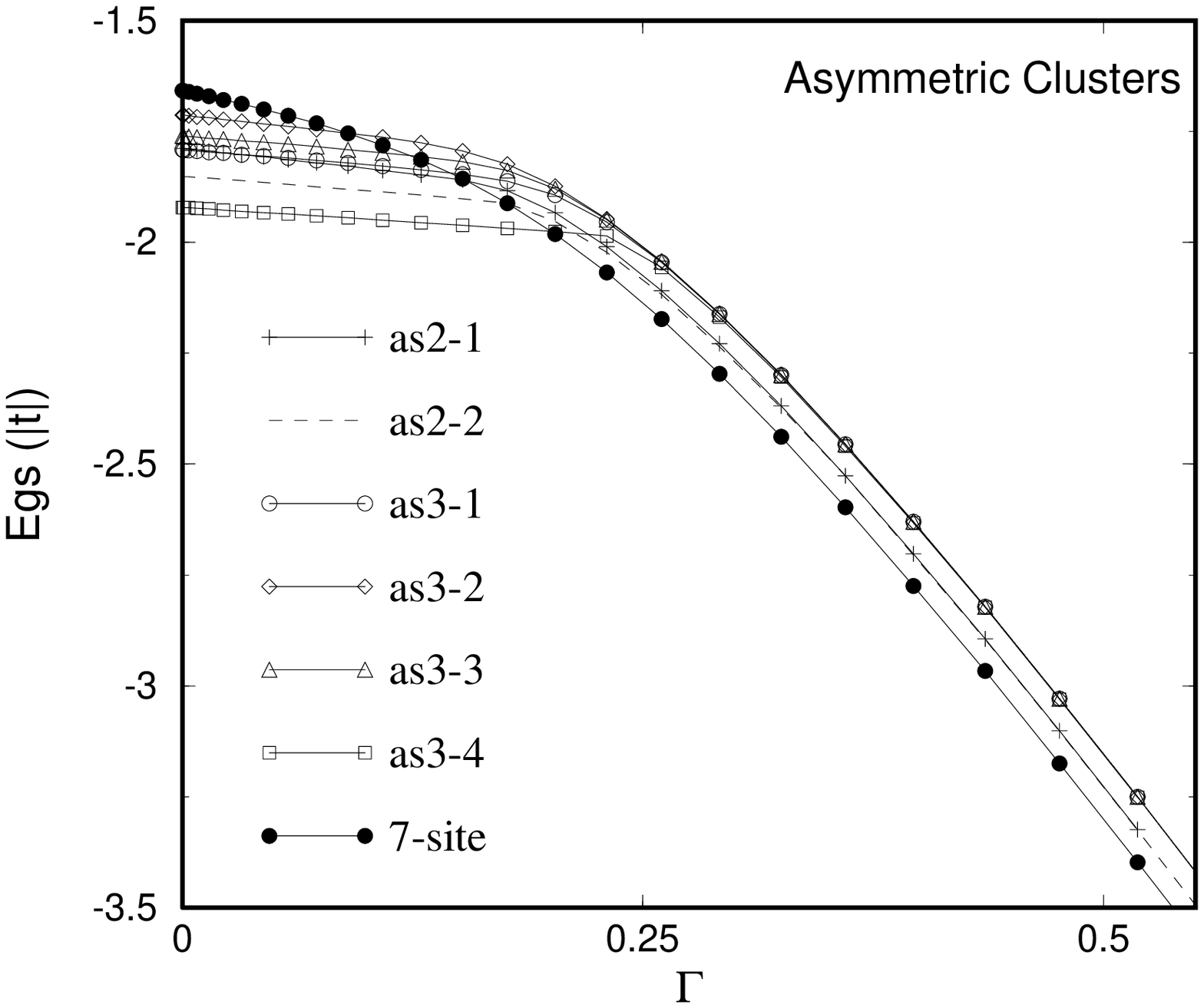,height=3.4in,width=3.4in,angle=0}}
\caption{Numerical results for ground state energy($E_{gs}$) of asymmetric
clusters}
\label{fig:asf1}
\end{figure}

\begin{figure}[htb]
\centerline{\psfig{figure=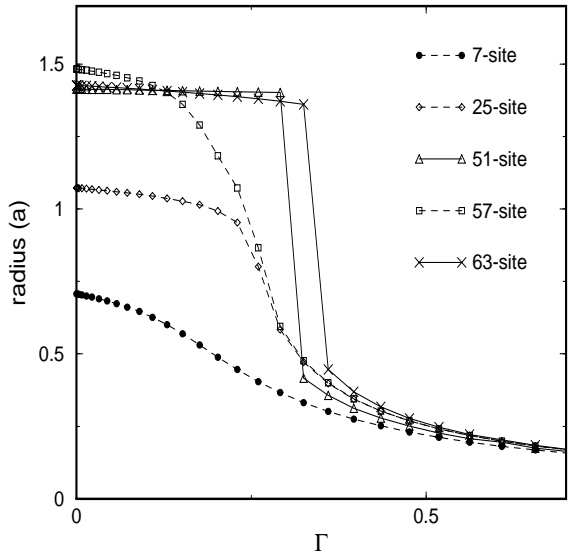,height=3.4in,width=3.4in,angle=0}}
\caption{Numerical results for ground state radius of symmetric clusters.
The radius is measured in unit of lattice constant $a$(the Mn-Mn
separation distance), and is defined to be 
$[\sum_{i}(\vec{r}_i-<\vec{r}>)^2\psi_i^2]^{1/2}$, where 
the index i runs over all Mn sites inside the cluster, and 
$<\vec{r}>\equiv\sum_i\vec{r}_i\psi_i^2$.} 
\label{fig:radius}
\end{figure}

Now we consider the full Hamiltonian
${\cal H}={\cal H}_t+{\cal H}_J+{\cal H}_{\rm ep}+{\cal H}_{\rm L}$
for all clusters examined in Sec. IV. FM spin alignment facilitates
hopping and causes energy lowering from delocalization, therefore
encourages the polaron to grow.  However, as $\Gamma$ increases,
the JT splitting will become the dominating influence
on ground state energy.  Greater localization enhances the JT energy
lowering, which increases localization of the electron.  On the
other hand, ${\cal H}_J$ and ${\cal H}_{\rm L}$
serve as penalties for spin mis-alignment and lattice distortions.
Our numerical studies find the optimal resolution of the
competition between these effects.  The 7-site cluster
becomes favored from $\Gamma >0.18$, as shown in Fig. \ref{fig:syf2}.
and Fig. \ref{fig:asf1}.  For all clusters considered, when $\Gamma$
becomes large enough, the size of the ground
state wavefunction shrinks to become that of the 7-site polaron,
as shown in Fig. \ref{fig:radius}. Enlarging the size
of the FM cluster to enhance delocalization energy
is then disfavored by the strong lattice polaron effect.
It is also clear that for extremely large $\Gamma$, the 1-site
lattice polaron will be the only form of electron state which exists.

Our numerical calculation
predicts the lattice distortion pattern of the 7-site ground state.
For $\Gamma=0.25$, $t=0.75$ eV, 
and $K=27.2$ eV/$\AA^2$ (obtained from Raman scattering in 
LaMnO$_3$), the ground state shown in Fig. \ref{fig:7Mn} has oxygen
displacement parameters $q_{11} \simeq 1.1$, $q_{13} \simeq 0.78$ and
$q_{12}=0$, which gives an outward displacement of $0.38\AA$ in 
the $\hat{z}$ direction, and smaller outward displacements of $0.036\AA$ 
in both $\hat{x}$ and $\hat{y}$ directions, for the 6 oxygens
surrounding the central Mn.  Displacements of oxygens
in the intermediate and outer layers are at least 10 times smaller.
As predicted perturbatively, $t_{1u}$, $t_{2u}$, 
$a_{2g}$, $q_{42}$ and $q_{43}$ distortions are absent. 


\section{Discussion}

When doping $x$ is non-zero, one should ask whether localized polaron
solutions will distribute homogeneously or will attract, causing
phase separation \cite{Moreo}.  
We have not addressed this issue, which requires
a more complicated calculation with additional Coulomb parameters
in the Hamiltonian.  Experiment is consistent with
a concentration interval up to $x=0.08$ where polarons are 
homogeneously distributed.  Our model shows that by this concentration,
randomly-distributed 7-site polarons will overlap significantly.
However, our model also shows that at concentrations
$x > 0.045$, polarons should be unstable relative to an
undistorted ground state with FM spin order and metallic conduction
by the doped electrons.  This effect is not seen in experiments.
Apparently alternate ground states, possibly involving organization
of polaronic distortions, occur and enable the system to remain 
non-metallic.

Without additional physics (such as defects), our model cannot
account for the observation \cite{Neum} that La concentrations
$x$ less than 0.02 yield less excess moment than expected from
7-site polarons.

Our model describes the
competing spin and lattice polaron effects.  We believe
that it contains the main features needed to describe
the system.  A test would be measurement of the
oxygen displacements which our model predicts.
The model omits non-adiabatic phonon effects,
spin quantization, temperature, and
polaron-polaron interactions\cite{Batista}.  For higher
doping levels or $T>0$, these may have a larger influence
and present challenges which could be worth pursuing if
experimental guidance improves.

\section*{Acknowledgment}
We thank J. J. Neumeier for suggesting this investigation;
J. J. Neumeier and J. L. Cohn for comments on the manuscript, and
V. Perebeinos and A. Abanov for discussions. This work was supported by NSF
grant No. DMR-0089492.


\begin{references}
\bibitem{note}	There is a factor of 2 change in the normalization
	of the displacement variables $Q\propto u_a-u_b$ in the present
	paper, relative
        to the convention used in Ref. \cite{Allen}.  For this reason,
        even though the coupling constant $g$ is exactly the same, the
	Hamiltonians Eqn.(\ref{eq:Hep},\ref{eq:HJT},\ref{eq:Hbr})
	look larger by a factor of 2.  When expressed in terms of the single
	atom displacements $u$, the Hamiltions are the same.
\end{references}
\end{document}